\documentclass[conference, 10pt]{IEEEtran}

\usepackage{amsmath}
\usepackage{amsfonts}
\usepackage{amssymb}
\usepackage{amsthm}
\usepackage{color}
\usepackage{array}
\usepackage{cite}
\usepackage{enumerate}
\usepackage{graphicx}
\usepackage[hang]{subfigure}
\usepackage{epstopdf}
\usepackage{epsfig}
\usepackage{footmisc}
\usepackage{amsbsy}
\usepackage{bm}
\usepackage{cases}
\usepackage{multirow}
\usepackage{algorithmic}
\usepackage{comment}
\usepackage{times}
\usepackage{paralist}
\usepackage[ruled]{algorithm2e}
\usepackage{setspace}
\usepackage[breakable]{tcolorbox}
\IEEEoverridecommandlockouts
\begin{document}
%
\title{Start Making Sense:\\ Semantic Plane Filtering and Control for \\ Post-5G Connectivity}
%
%
%

\author{Petar Popovski~\IEEEmembership{Fellow,~IEEE,}
        and Osvaldo Simeone~\IEEEmembership{Fellow,~IEEE}
\thanks{Authors in alphabetical order. Petar Popovski (petarp@es.aau.dk) is with the Department of Electronic Systems, Aalborg University, Aalborg, Denmark. Osvaldo Simeone (osvaldo.simeone@kcl.ac.uk) is with the Centre for Telecommunications Research, King’s College London, London, United Kingdom. This work has received funding from the European Research Council
(ERC) under the European Union Horizon 2020 research and innovation program (grant agreements 725731 and 648382). }}
\maketitle



%
\IEEEpeerreviewmaketitle

%
%
%
%

\vspace{-1.8cm}
\
\begin{quote} \emph{You're talking a lot, but you're not saying anything}\\
\emph{When I have nothing to say, my lips are sealed} \\
\emph{Say something once, why say it again?} \\
Psycho Killer, Talking Heads \end{quote}

\vspace{1cm}

\section{Context}

This short position paper puts forth a proposal for an protocol-level solution for the post-5G era that is based on the following current trends.
\begin{itemize}
    \item In the post-5G era, solutions for wireless connectivity will be abundant and mature, shifting the performance bottleneck to the core network segment. While this does not imply that no further advances are expected in physical-layer technology, it is envisaged there will be more substantial and conceptual developments in layers and services that leverage physical-layer connectivity primitives. 
    \item Recent advances in Machine Learning (ML), also popularly referred to as Artificial Intelligence (AI), will enable the extraction of information by means of pattern recognition within complex data streams at all layers of the protocol stack, even when the information of interest is not explicitly encoded (see, e.g., \cite{simeone2018very}).
   \item With the convergence of heterogeneous data- and task-oriented services on the cellular infrastructure, data streams will contain a large fraction of \emph{semantic overhead}, that is, of data that is delivered to the application layer, but ends up not being relevant or useful. A notable example is the transmission of data for the training of ML/AI models that does not lead to significant updates in the current model or whose source cannot be trusted (see, e.g., \cite{chen2018lag}). This issue is a direct consequence of Claude Shannon's dominant design paradigm, according to which the network is largely oblivious to the semantics of the information bits being transported~\cite{shannon1948mathematical}.
   \item 5G networks will be characterized by an increased \emph{protocol overhead} in order to ensure security, privacy, and provenance, including through access to Distributed Ledger Technology (DLT)~\cite{christidis2016blockchains}.
\end{itemize}

\begin{figure}[!t]
\centering
\includegraphics{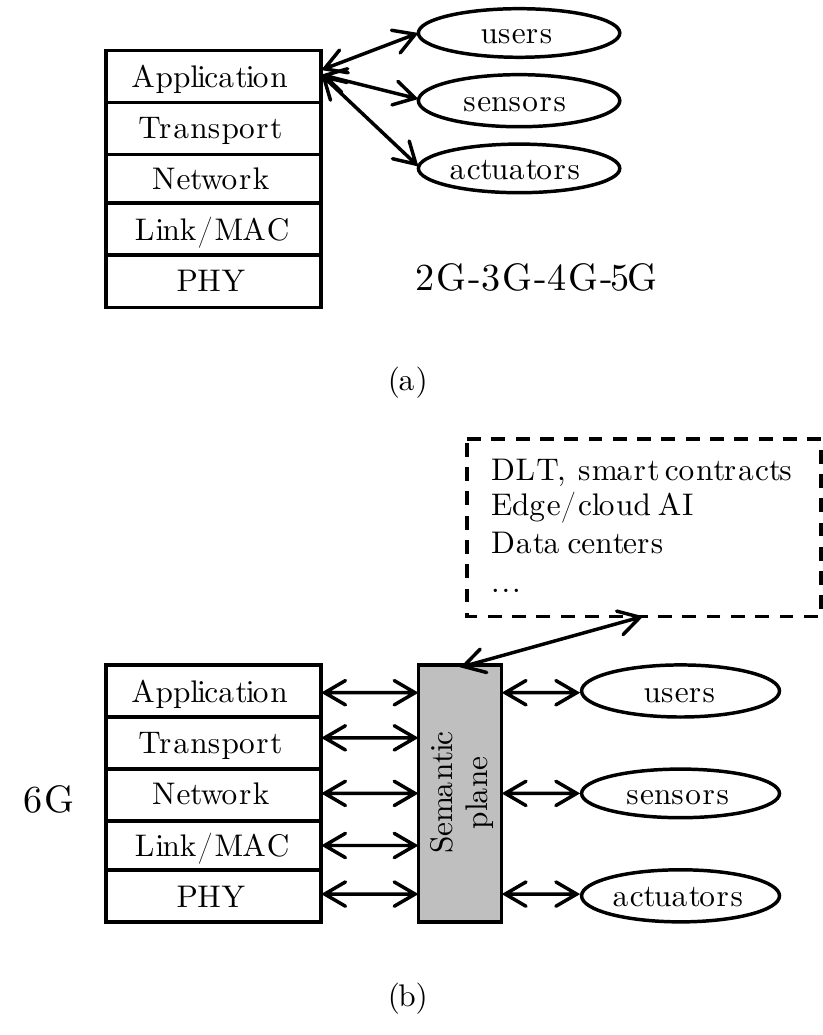}%
\label{fig1}
\caption{Proposed architectural evolution.}
\end{figure}



\section{Semantic-Plane Filtering and Control}

In light of the above trends, as illustrated in Fig. 1, we propose to move away from the current protocol architecture whereby the semantic interface between users, sensors, and actuators is limited to the application layer. Instead, in the proposed architecture, the protocol stack is augmented with a \emph{semantic plane} that provides a common interface between, on the one end, users, sensors, and actuators, and, on the other end, all layers of the protocol stack. The semantic plane carries out signal processing and filtering of information at all layers of the protocol stack by means of its access to edge/ cloud intelligence, data centers, as well as DLT in the form of blockchains or smart contract ledgers. The semantic plane also controls information flow between layers through a well-defined software interface. 

The semantic plane extracts information from data streams at all layers of the protocol by means of ML/AI pattern recognition algorithms implemented at edge or core cloud platforms. The extracted information can be used to: \begin{itemize}
\item\emph{filter} irrelevant or untrusted streams by cross-checking against the state of the ML/AI models or trusted ledgers;
\item \emph{control} the operation of the communication protocol; or
\item \emph{control} connected actuators and sensors.   
\end{itemize}

As a result of semantic filtering, both semantic and protocol overheads can be potentially significantly reduced, as illustrated in Fig. 2. Furthermore, semantic control of all layers of the protocol stack can enable a more efficient use of communication and computing resources and potentially allow the introduction of new applications.

\section{Examples of the Semantic Plane Functionality}

Specific examples of the applications of the proposed semantic plane include the following:
\begin{itemize}
\item \emph{Physical-layer provenance filtering}: Validate data provenance using radio fingerprinting or location-based authentication through baseband signal processing (see, e.g., \cite{xu2016device}). This reduces protocol overhead by removing the need for higher-layer operations, and it also reduces semantic overhead by filtering out insecure links.
\item \emph{Physical-layer computing}: Compute aggregated statistics by means of physical-layer computing, hence reducing semantic and protocol overheads by avoiding the need to transmit separate data points \cite{nazer2011compute, compfed18, 2019arXiv190100844A}.
\item \emph{Physical-layer remote radio control}: Allow access to physical resources by means of smart locks accessed via radio signatures controlled through smart contracts. For example, once rent is paid, a smart contract controls the physical layer transmission of a radio signature to open a smart lock \cite{roush2018twelve}. This results in reduced protocol overhead and it enables new applications.
\item \emph{Physical-layer integration of mmwave/THz radar and communication}: Use the same radio interface for both radar and communications~\cite{choi2016millimeter}, hence increasing efficiency and enabling new applications such as gesture-based interfaces. 
\item \emph{MAC-layer retransmission control}: Retransmit only data that is expected to be still relevant on the basis of the internal state of ML/AI models (see, e.g., \cite{zhu2018towards}), hence reducing protocol and semantic overhead.
\item \emph{Network-layer traffic-based routing}: Classify traffic on the basis of its network-level traces (see, e.g., \cite{nguyen2008survey}) and configure routers accordingly, hence reducing protocol overhead.
\item \emph{Transport-layer semantic-based access and congestion control}: Inject data in the network only when relevant and trusted, hence reducing semantic overhead. A notable example is given by sensors and actuators that operate on the basis of input from local or cloud-based predictive models. Accordingly, based on past data collected by sensors and actuators, an ML/AI module can instruct a sensor on how to carry out sampling and transmission and can inform an actuator of any change in its control module only when the local model is outdated. 

\begin{figure}[!t]
\centering
\includegraphics{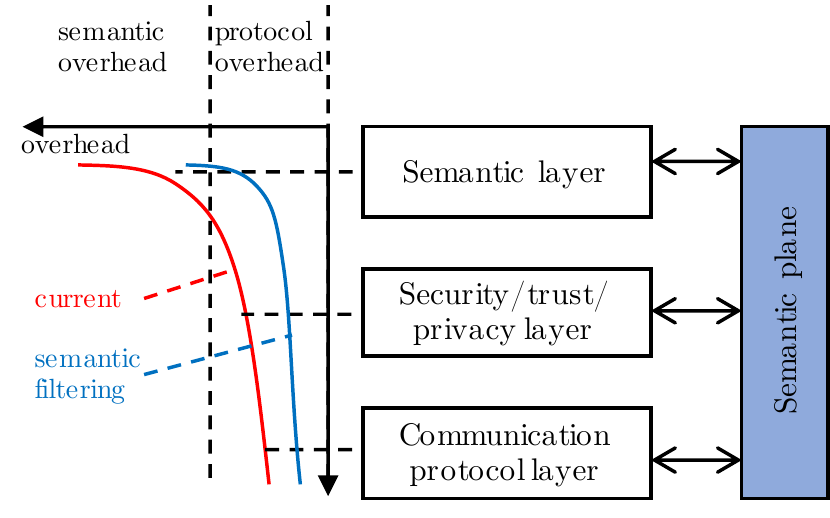}%
\label{fig2}
\caption{The role of semantic filtering in decreasing both semantic and protocol overheads.}
\end{figure}

\item \emph{Application-level aggregation for DLT transactions}: Aggregate transactions from multiple applications running on a device or a network node in a single verifiable unit (e.g., block) prior to communications within a distributed ledger for, e.g., financial transactions or smart contracts, hence reducing the protocol overhead.

\item \emph{Intent-based networking}: Upon receiving users' instructions via natural language or visual interfaces, the network automatically reconfigure itself at all layers of the protocol stack in order to carry out the described task in the most efficient manner.
\end{itemize}

\section{Outlook}

A standardization of the semantic plane architecture would by and large revolve around the definition of effective interfaces between communication layers and application programming interfaces towards users, actuators, and devices. A successful definition would provide enough flexibility to enable applications such as those listed above without causing excessive inefficiencies that would offset the potential gains discussed above. 
The availability of well-defined software interfaces for semantic filtering and control could in fact invalidate the current ``next-G'' paradigm for mobile wireless evolution and standardization, ushering in an era of continuous ``open-source'' improvements and extensions.

\ifCLASSOPTIONcaptionsoff
  \newpage
\fi



%

\bibliographystyle{IEEEtran}
\bibliography{IEEEabrv,references}

%




\end{document}